\documentclass[11pt,a4paper,notitle]{article}
\usepackage{amsmath,amsfonts,amssymb,graphicx,lmodern}
\usepackage[left=2.5cm,right=2.5cm,top=2.5cm,bottom=2.5cm]{geometry}

\usepackage{subfigure,color,float}
\usepackage{mathptmx}

\usepackage[english]{babel}
\usepackage[round,authoryear]{natbib}
\makeatletter
\renewcommand{\@seccntformat}[1]{\csname the#1\endcsname.\quad}
\makeatother

\usepackage{titlesec}
\titleformat*{\section}{ \fontsize{11}{11}\bfseries}
\titleformat*{\subsection}{\fontsize{11}{11}\bfseries}

\begin{document}

{\centering \fontsize{14}{14}\selectfont \bf Connection between Dark Matter Abundance and  Primordial Tensor Perturbations \\}
\bigskip
{\centering \fontsize{10}{10}\selectfont P. S. Bhupal Dev$^1$, Anupam Mazumdar$^{2}$, Saleh Qutub$^{2,3}$ \\
\fontsize{10}{10}\selectfont $^1$~Consortium for Fundamental Physics, School of Physics and Astronomy, \\
\fontsize{10}{10}\selectfont University of Manchester, Manchester, M13 9PL, United Kingdom.\\
\fontsize{10}{10}\selectfont $^2$~Consortium for Fundamental Physics, Physics Department, \\ Lancaster University, LA1 4YB, United Kingdom.\\
\fontsize{10}{10}\selectfont $^{3}$~Department of Astronomy, King Abdulaziz University, Jeddah 21589, Saudi Arabia.\\}

\bigskip
\bigskip

{\noindent \bf ABSTRACT}

Primordial inflation and Dark Matter (DM) could both belong to the hidden sector. It is therefore plausible that the inflaton, which drives inflation, 
could couple to the DM either directly or  indirectly, thus providing a common origin for both luminous and non-luminous matter. We explore this interesting 
possibility and show that, in certain scenarios, the DM mass can be correlated with the tensor-to-scalar ratio. This correlation might 
provide us with a window of opportunity for unravelling the properties of DM beyond the standard freeze-out paradigm.

\bigskip


%%%%%%%%%%%%%%%%%%%%%%%%%%%%%%%%%%%%%%%%%%%%%%%%%%%%%%%%%%%%%%%%%%%%%%%%%%%%%%%%%%%%%%%%%%%%%%%%%%%%%%%%%%%%%%%%%%%%%%%%%%%%%%
%%%%%%%%%%%%%%%%%%%%%%%%%%%%%%%%%%%%%%%%%%%%%%%%%%%%%%%%%%%%%%%%%%%%%%%%%%%%%%%%%%%%%%%%%%%%%%%%%%%%%%%%%%%%%%%%%%%%%%%%%%%%%%
\section{INTRODUCTION}
\noindent

\begin{figure}[ht] 
\centering
\subfigure[]{\label{fig:decay}
\includegraphics[width=2.7cm]{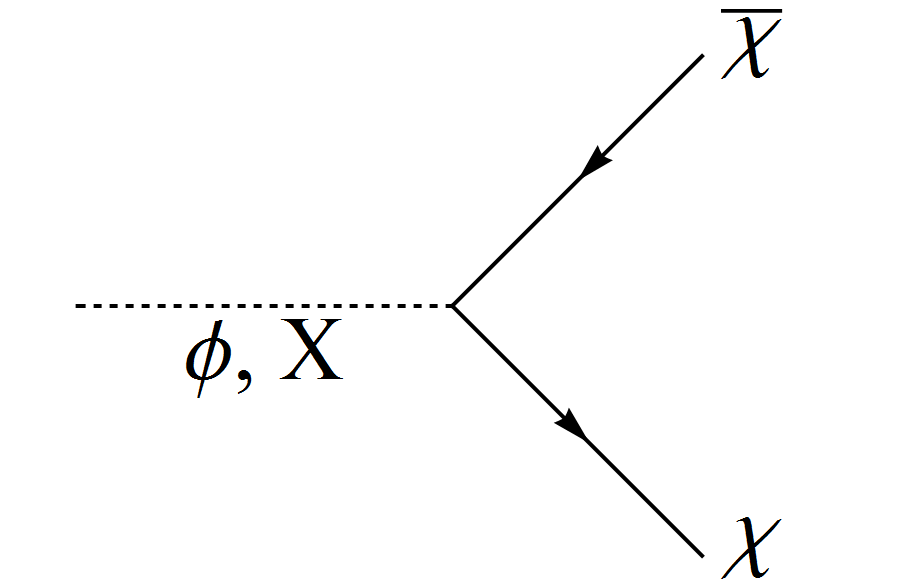}}
\hspace{2 cm}
\subfigure[]{\label{fig:annihilation}
\includegraphics[width=3.5cm]{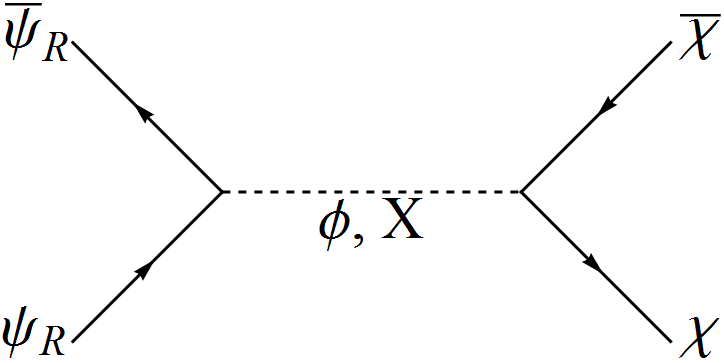}}
\caption{DM production from (a) non-thermal decay of inflaton/heavy scalar and (b) thermal scatterings with the SM d.o.f mediated by inflaton/heavy scalar field.}
\label{fig:feyndiag}
\end{figure}

Primordial inflation is one of the simplest paradigms to explain the formation of 
large scale structures in our Universe and the observed features of the Cosmic Microwave Background spectrum~\citep{Ade:2015xua}. Inflation is driven by the vacuum energy density of a scalar field, known as the inflaton, whose origin usually requires some beyond the Standard Model (SM) physics~\citep{Mazumdar:2010sa}. On the other hand, various astrophysical and cosmological observations~\citep{Bertone:2004pz} strongly suggest the existence of  a 
non-luminous, non-baryonic form of matter, known as Dark Matter (DM). 
%, which follows a universal density profile as an essential ingredient 
%for the hierarchical structure formation~\citep{Bertone:2004pz}, as suggested by detailed $N$-body simulations~\citep{Navarro:1995iw}, and semi-analytics~\citep{Dayal:2014cda}.
Although the masses and interactions of either inflaton or DM are still unknown, 
both must couple to the SM in some way, if not directly. 
Technically speaking, they can be considered to be SM gauge singlets, and therefore, could both belong to 
the dark or the hidden sector. In this case, one can imagine that the net DM, which we assume to be cosmologically stable, can be created via two processes:

\noindent
{\bf (a) Decay}: The inflaton $\phi$ could directly couple to the DM $\chi$ via renormalizable interactions, as shown in Fig.~\ref{fig:decay}. 
For concreteness, we assume a fermionic DM, so that the interaction is of the Yukawa type, i.e. $\phi\overline{\chi}\chi$. We could also have a
scalar DM with a trilinear coupling to inflaton. Apart from the inflaton itself, any heavy hidden sector scalar field $X$ could also directly decay to DM via renormalizable interactions.

\noindent
{\bf (b) Scattering}: Inflaton must couple to the SM degrees of freedom (d.o.f) for the success of Big Bang Nucleosynthesis. Since the SM fermions are chiral, a SM singlet inflaton can in principle couple to the right-handed (RH) fermions $\psi_R$ via renormalizable interactions of the form $\phi \overline{\psi}_R \psi_R$. Therefore, we can also create DM via {\em inflaton mediation}, as shown in Fig.~\ref{fig:annihilation}. One can generalize this scenario to envisage that  any heavy scalar mediator  $X$ could connect the SM d.o.f with the dark sector.  

%%%%%%%%%%%%%%%%%%%%%%%%%%%%%%%%%%%%%%%%%%%%%%%%%%%%%%%%%%%%%%%%%%%%%%%%%%%%%%%%%%%%%%%%%%%%
%%%%%%%%%%%%%%%%%%%%%%%%%%%%%%%%%%%%%%%%%%%%%%%%%%%%%%%%%%%%%%%%%%%%%%%%%%%%%%%%%%%%%%%%%%%%

A rather natural outcome of this simple scenario is that the scale of inflation, determined by the inflaton potential $V(\phi)$, can be correlated with the DM properties in a rather intriguing way. If, for some reason, the DM does not fully thermalize with the primordial plasma during its evolution, it can in principle retain the memory of how it was excited at the first instance, either (a) directly via the inflaton decay or (b) indirectly via scatterings mediated by the inflaton or a heavy scalar field. In case (a), the DM is essentially decoupled from the thermal bath since its creation. This leads to a non-thermal DM scenario, where the DM relic abundance is directly determined by the initial inflaton energy density~\citep{Allahverdi:2002nb, Allahverdi:2002pu}. In case (b), if the effective coupling of the DM to the SM d.o.f is too small to fully thermalize the DM with the bath, but sufficient enough to produce the observed abundance of DM, this leads to the Feebly Interacting Massive Particle (FIMP) or freeze-in DM scenario~\citep{Hall:2009bx}. In both cases, the final DM relic density is sensitive to the initial conditions set by inflation~\citep{Dev:2013yza}, unlike in the standard thermal weakly interacting massive particle (WIMP) scenario~\citep{Kolb:1990vq}, thereby providing the unique possibility to directly link the DM properties with inflation. 

In this paper, we show that in the non-WIMP scenarios, it is indeed possible to establish a connection between the DM and inflaton sectors via the scale of inflation, which can be determined by measuring
the primordial tensor-to-scalar ratio.
Note that for the WIMP DM scenario with a relatively large DM coupling to the SM d.o.f, there are many observational constraints from direct/indirect searches~\citep{Bertone:2004pz}, but few are applicable to the non-standard DM scenarios (a) and (b) discussed above. Therefore, the window of opportunity established in this paper is extremely useful for probing such DM candidates. 
We illustrate this for a simple class of inflationary models with an $\alpha$-attractor potential~\citep{DeFelice:2011jm, Kallosh:2013yoa}, but our results could be easily extended to other inflationary potentials.  
%with $\alpha$ being a free parameter that determine the shape of the potential. For $\alpha \rightarrow \infty$, the inflationary potential mimics the shape of the quadratic one around the minimum,
%whereas for $\alpha \sim {\cal O}(1)$, the inflaton potential around the minimum is not generally quadratic. However for $\phi \ll M_P$, it can be approximated by a quadratic one albeit being steeper than that for $\alpha \gg 1$.

%%%%%%%%%%%%%%%%%%%%%%%%%%%%%%%%%%%%%%%%%%%%%%%%%%%%%%%%%%%%%%%%%%%%%%%%%%%%%%%%%%%%%%%%%%%%%%%%%%%%%%%%%%%%%%%%%%%%%%%%%%%%%%
\section{A BRIEF REVIEW ON INFLATIONARY SET UP}
\indent

Typically, the scale of inflation can be observationally determined by the tensor-to-scalar ratio $r= {\cal P}_T/{\cal P}_S$, where
${\cal P}_S$ is the amplitude of the scalar perturbations given by %${\cal P}_s= 2.196_{-0.060}^{+0.051} \times 10^{-9}$ 
${\cal P}_s= 2.142_{-0.049}^{+0.049} \times 10^{-9}$~\citep{Ade:2015xua} and ${\cal P}_T$ is the amplitude of the tensor power spectrum.
 The Hubble rate $H=V(\phi)/3M_P^2$, where $M_P=2.4\times 10^{18}$ GeV is the reduced Planck mass, is given by~\citep{Kolb:1990vq}
\begin{eqnarray}\label{SOI}
H \ \simeq \  3\times 10^{-5} \left(\frac{r}{0.1}\right)^{1/2}M_P\, .
\end{eqnarray}
Thus, it is possible that by measuring $r$ one might get some insight about the DM relic abundance using the production mechanisms given in Fig.~\ref{fig:feyndiag}. In order to illustrate this, let us consider a class of inflationary models given by the following potential~\citep{DeFelice:2011jm, Kallosh:2013yoa}:
\begin{eqnarray}
V(\phi) &=& \frac{3}{4} M_P^2 M^2
\left( 1-e^{-\sqrt{\frac{2}{3\alpha}}
\frac{\phi}{M_P}} \right)^2 \ \equiv \  \frac{3}{4} M_P^2 M^2 \left[1- x(\phi) \right]^2,
\label{eq:pot}
\end{eqnarray}
where $\alpha$ is a free parameter and $M$ is some mass scale governing inflation. 
%
%%%%%%%%%%%%%%%%%%%%%%%%%%%%%%%%%%%%%%%%%%%%%%%%%%%%%%%%%%%%%%%%%%%%%%%%%%%%%%%%%%%%%%%%%%%%
%%%%%%%%%%%%%%%%%%%%%%%%%%%%%%%%%%%%%%%%%%%%%%%%%%%%%%%%%%%%%%%%%%%%%%%%%%%%%%%%%%%%%%%%%%%%
%
For $\alpha=1$, this is just the Starobinsky model~\citep{Starobinsky:1980te}, whereas in the limit $\alpha\rightarrow \infty$, Eq.~\eqref{eq:pot} reduces to the simple quadratic chaotic inflation potential~\citep{Linde:1983gd} with constant mass, $ M/\sqrt{\alpha}$. In this class of models, inflation occurs above the scale of $M_P$ and terminates at
\begin{eqnarray} \label{eq:phiend}
\phi_{\rm end} \ \simeq \ \sqrt{ \frac{3\alpha}{2} } \ln \left( 1+\frac{2}{\sqrt{3\alpha}} \right) \ M_P \, .
\label{eq:pend}
\end{eqnarray}
In the limit $\alpha\rightarrow \infty$, $\phi_{\rm end} \rightarrow \sqrt{2} M_P$. The  potential at the end of inflation is then given by
\begin{equation} \label{eq:Vend}
V_{\rm end} \ \equiv \ V(\phi_{\rm end}) \ \simeq \ \frac{9 M^2 M_P^2}{(2 \sqrt{3} + 3 \sqrt{\alpha})^2}
\, .
\end{equation}
The mass scale $M$ can be expressed in terms of ${\cal P}_s$~\citep{Kallosh:2013yoa,Ozkan:2015iva}
\begin{eqnarray}
\frac{M}{M_P} \ = \ \sqrt{ \frac{128 \pi^2 {\cal P}_s}{3 \alpha} } \frac{x_{\rm obs}}{(1-x_{\rm obs})^2}
\, ,
\label{eq:M_Ps}
\end{eqnarray}
where $x_{\rm obs} \equiv x(\phi_{\rm obs})$ which can be evaluated once the number of e-foldings $N$ is known~\citep{Kallosh:2013yoa,Ozkan:2015iva}: 
\begin{eqnarray}
N \ = \ \frac{3}{4}\alpha \left( \frac{1}{x_{\rm obs}}-\frac{1}{x_{\rm end}} \right)
+\left( \frac{3}{4} \alpha-\frac{1}{2} \right) \ln \left(
\frac{x_{\rm obs}}{x_{\rm end}} \right)
\, ,
\label{eq:N}
\end{eqnarray}
where $x_{\rm end}\equiv x(\phi_{\rm end}) = (1 + 2/\sqrt{3 \alpha})^{-1} $. 

Using Eq.~(\ref{eq:N}), the tensor-to-scalar ratio for the potential in Eq.~(\ref{eq:pot}) is given by
\begin{eqnarray}
r (\alpha) \ = \ \frac{64x^2_{\rm obs}(\alpha)}{3\alpha[1-x_{\rm obs}(\alpha)]^2}
\, .
\label{eq:r}
\end{eqnarray}
%
%which can be also evaluated with the help of  once $N$ is known. 
For $\alpha = {\cal O}(1)$, $M \simeq \sqrt{24 \pi^2 \alpha {\cal P}_s} \ M_P/N $ and $r \simeq 12 \alpha/N^2 $, whereas for $\alpha \gg 1$, $M \simeq \sqrt{6  \pi^2 \alpha {\cal P}_s} \ M_P/N $ and $r \simeq 8/N $. 
\begin{figure}[t] 
\centering
\includegraphics[width=5.9cm]{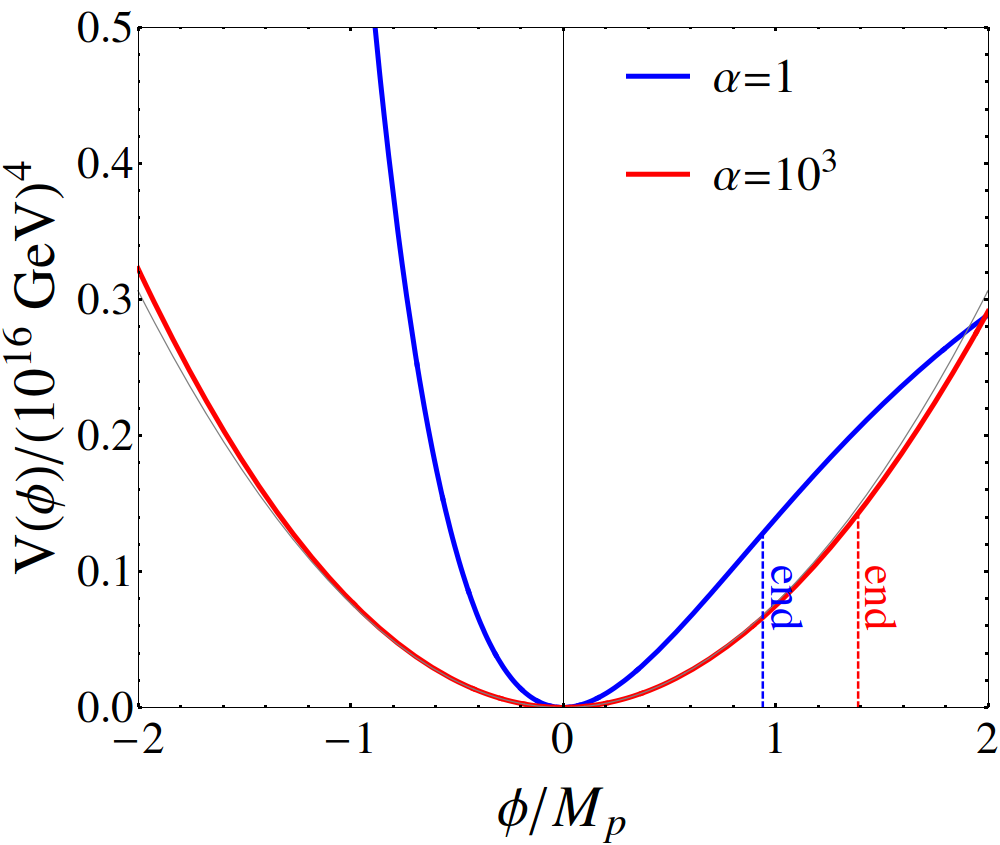}
\caption{Shape of the potential given by Eq.~(\ref{eq:pot}) for $\alpha =1$ and $10^3$. Here `end' refers to the end of inflation, as given by Eq.~\eqref{eq:pend}. The gray curve (below the red curve) denotes a quadratic potential. 
}
\label{fig:pot}
\end{figure}
In Fig.~\ref{fig:pot} we show a plot of the potential Eq.~(\ref{eq:pot}) for $\alpha =1$~which leads to $r\simeq {\cal O}(0.004)$, and $\alpha =10^3$~which leads to $r\simeq {\cal O}(0.1)$. We also show the value of $\phi$ at which inflation terminates in each case. 
%
%%%%%%%%%%%%%%%%%%%%%%%%%%%%%%%%%%%%%%%%%%%%%%%%%%%%%%%%%%%%%%%%%%%%%%%%%%%%%%%%%%%%%%%%%%%%

%%%%%%%%%%%%%%%%%%%%%%%%%%%%%%%%%%%%%%%%%%%%%%%%%%%%%%%%%%%%%%%%%%%%%%%%%%%%%%%%%%%%%%%%%%%%
%

\begin{figure}[t] 
\centering
%\subfigure[]{\label{fig:Vend-alpha}
%\includegraphics[width=6.0cm]{Vend-alpha.png}}
%\hspace{0cm}
\subfigure[]{\label{fig:Vend-r}
\includegraphics[width=6.1cm]{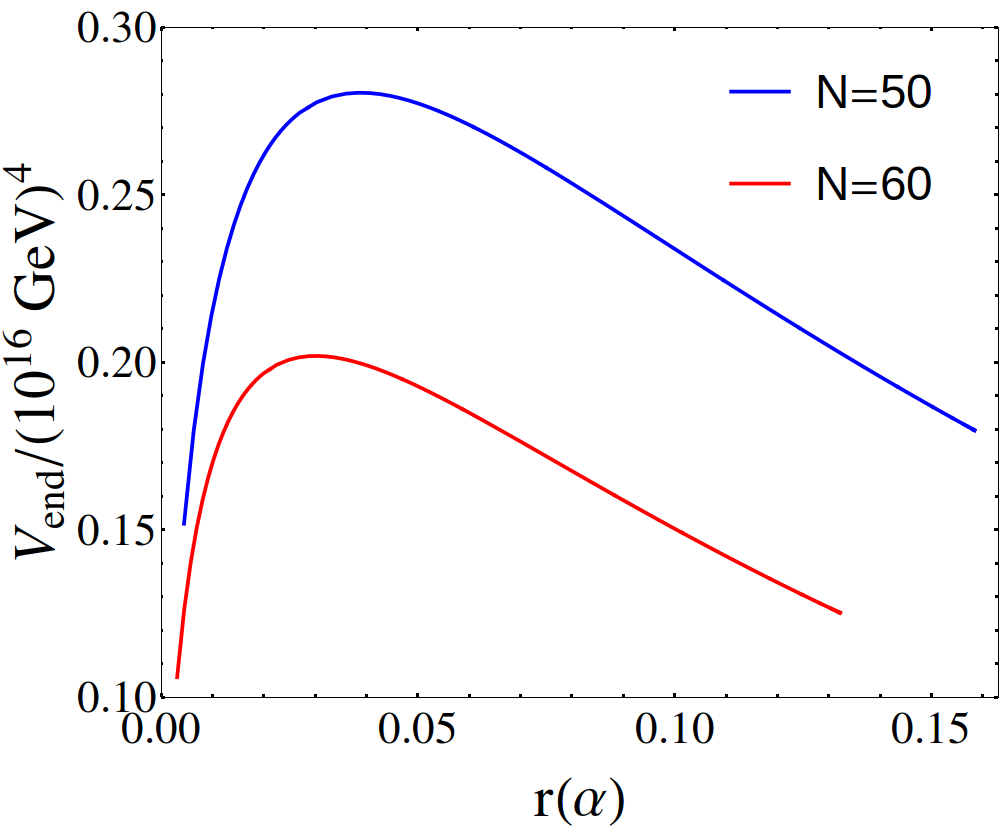}}
\hspace{0cm}
\subfigure[]{\label{fig:mphi-r}
\includegraphics[width=6.0cm]{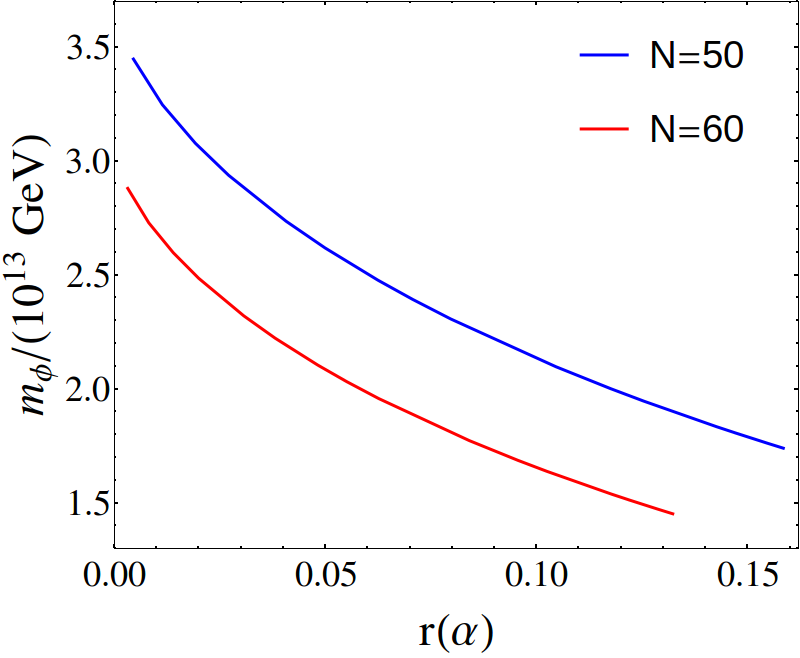}}
\caption{(a) The inflaton energy density at the end of inflation and (b) effective inflaton mass for $\phi\ll M_P$ as a function of $r(\alpha)$ for $N=50~(60)$ shown by the blue (red) curve.} 
\label{fig:Vend-mphi}
\end{figure}

In Fig.~\ref{fig:Vend-r}, we show the inflaton potential energy at the end of inflation as a function 
%of $\alpha$ (left panel)  and as a function 
of $r(\alpha)$ 
%(right panel)
for 50 and 60 e-foldings, since the precise value of $N$ depends on the details of reheating process~\citep{Lyth:2009zz}.
%Assuming the Universe behaves as a matter dominated one during the reheating phase, where inflaton decay takes place, the number of e-foldings, $N$, can be estimated as~\citep{Lyth:2009zz}
%
%\begin{equation}
%N \simeq 56 - \frac{2}{3} {\rm ln} \left( \frac{10^{16} {\rm GeV}}{V_{\rm end}^{1/4}} \right) - \frac{1}{3} {\rm ln} \left( \frac{10^{9} {\rm GeV}}{T_{\rm rh}} \right)
%\, ,
%\label{eq:N2}
%\end{equation}
%
%where $T_{\rm rh}$ denotes the reheating temperature.
Once inflation ends, the field $\phi$ starts oscillating around the minimum of the potential with an effective mass
\begin{eqnarray}
m_{\phi, \rm eff} \ \simeq \ \left(\frac{\partial^2 V(\phi)}{\partial \phi^2} \right)^{1/2} \ \simeq \  \frac{M}{\sqrt{\alpha}} \left( 2  e^{-2 z} - e^{-z}  \right)^{1/2} \, ,
\end{eqnarray}
where $z = \sqrt{2/3\alpha} ~ \phi/M_P$. The evolution of $\phi$ is governed by the following background equation of motion during the inflaton oscillations:
\begin{eqnarray}
\frac{\partial^2\phi}{\partial t^2} + (3 H +\Gamma_\phi)\frac{\partial \phi}{\partial t} + \frac{\partial V(\phi)}{\partial \phi} \ \simeq \ 0\,,
\label{eq:phi}
\end{eqnarray}
where $\Gamma_\phi$ is the decay rate of the 
inflaton field. Eq.~\eqref{eq:phi} holds true as long as  
$m_\phi \gg H(t)$, so that the thermal and backreaction effects can be neglected. For the potential given by Eq.~\eqref{eq:pot},  the amplitude of $\phi$ oscillation quickly drops with the expansion of the Universe right after the end of inflation; it becomes less than 0.2 of its initial value after only one oscillation. Thus, after a few oscillations, $\phi$ becomes confined to a small region around the minimum of the potential for which the potential can be approximated by a quadratic one with a constant mass, $m_\phi \equiv m_{\phi, \rm eff}(\phi \ll M_P)  \simeq  M/\sqrt{\alpha}$.
Note that in the limit $\alpha \rightarrow \infty$, $m_{\phi, \rm eff}$ globally approaches $M/\sqrt{\alpha} \simeq \sqrt{6  \pi^2 {\cal P}_s} \ M_P/N$. In Fig.~\ref{fig:mphi-r}, we show $m_\phi$ as a function of $r$ for $N = 50$ and 60. 

%%%%%%%%%%%%%%%%%%%%%%%%%%%%%%%%%%%%%%%%%%%%%%%%%%%%%%%%%%%%%%%%%%%%%%%%%%%%%%%%%%%%%%%%%%%%%%%%%%%%%%%%%%%%%%%%%%%%%%%%%%
\section{DARK MATTER PRODUCTION FROM INFLATON DECAY PRODUCTS}
\indent

The renormalizable interactions involving the inflaton $\phi$, DM $\chi$, heavy mediator $X$ and the relevant SM fields (RH fermions $\psi_R$ and Higgs doublet $\varphi$) are given by
\begin{eqnarray}\label{interaction}
-\mathcal{L}_{\rm int}  \ \supset  \  y_{\!_{\phi \psi} } \phi \overline{\psi}_R \psi_R \: + \: \frac{1}{2} y_{\!_{\phi \varphi}}^2 \phi^2 \varphi^\dag \varphi \: + \: \frac{1}{2} y_{\!_{\phi X}}^2 \phi^2 X^2  \: + \: y_{\!_{\phi \chi}} \phi \overline{\chi} \chi  \: + \: y_{\!_{X \chi}} X \overline{\chi} \chi  \: + \: y_{\!_{X \psi} }X \overline{\psi}_R \psi_R  \: + \: {\rm H.c.} \, .
\label{eq:2}
\end{eqnarray}
Note that $\chi$ belonging to the hidden sector does not have a direct coupling to the SM fermions and the only indirect coupling arises by integrating out the mediator or the inflaton field in Eq.~(\ref{eq:2}).
Therefore, the effective interaction between the SM d.o.f and DM will be determined by a dimension-$6$, four-Fermion operator $\overline\chi \chi\overline\psi_R\psi_R/m_\phi^2$ or $ \overline\chi\chi\overline\psi_R\psi_R/m_X^2$, which will be respectively suppressed by the mass square of inflaton or the heavy mediator field, with ${\cal O}(1)$ Yukawa couplings. This naturally leads to a FIMP or non-thermal DM scenario which we are interested in here~\citep{Dev:2013yza,Blennow:2013jba,Baer:2014eja,Elahi:2014fsa}. We will keep our discussion general, without referring to any particular DM model.

In order to have a standard radiation-dominated era just after reheating, we require $ y_{\!_{\phi \chi}}, y_{\!_{\phi \varphi}} ,y_{\!_{\phi X}} \ll y_{\!_{\phi \psi}} $.
For the sake of simplicity and illustration, we may assume the couplings $y_{\!_{\phi \varphi}}, y_{\!_{\phi X}} \approx 0$. Any reasonable value of $ y_{\!_{\phi \varphi}} (y_{\!_{\phi X}})  \gtrsim 10^{-6}$ would in principle lead to a non-perturbative production of $X (\varphi)$~\citep{Shtanov:1994ce,Kofman:1994rk}, but in this case inflaton does not decay completely. One 
would still require the inflaton to decay perturbatively, which will be guaranteed to happen in our case via $y_{\!_{\phi \psi} } \phi \overline{\psi}_R \psi_R$~\citep{Allahverdi:2010xz}. Thanks to small Yukawa interactions, we can also ignore issues like fermionic preheating~\citep{Giudice:1999fb} or fragmentation of the inflaton~\citep{Enqvist:2002rj,Enqvist:2002si}. Further, we require that $ y_{\!_{\phi \psi}} \ll 1$ in order to avoid radiative corrections to the inflationary potential
~\citep{Coleman-Weinberg} and thermal corrections~\citep{Drewes:2014pfa}.

%~\footnote{Such radiative corrections cancel out in supersymmetric models due to the opposite sign contributions from the superpartners. However in such models, $ y_{\!_{\phi \psi}}$ has to be small enough such that $T_{\rm rh} \lesssim 10^9$~GeV in order to avoid overproduction of gravitinos.} and thermal corrections~\citep{Drewes:2014pfa}.

%%%%%%%%%%%%%%%%%%%%%%%%%%%%%%%%%%%%%%%%%%%%%%%%%%%%%%%%%%%%%%%%%%%%%%%%%%%%%%%%%%%%%%%%%%%%%%%%%%%%%%%%%%%%%%%%%%%%%%%%%%%%%%%
\subsection{DM Production}
\indent

Let us now compute the DM production rates due to both the channels shown in Fig.~\ref{fig:feyndiag}. Besides decaying to DM pairs, the inflaton will decay dominantly 
to the SM radiation and its total decay rate is given by 
\begin{eqnarray}
\Gamma_\phi  & \simeq &  \Gamma(\phi \rightarrow \overline{\psi} \psi) + \Gamma(\phi \rightarrow \overline{\chi} \chi) \nonumber \\
& \simeq & \sum_\psi c_\psi y^2_{\!_{\phi \psi}} \frac{ m_{\phi, \rm eff}}{8 \pi} \left( 1- \frac{4 m_{\psi, \rm eff}^2}{ m_{\phi, \rm eff}^2}\right)^{3/2}
+
\ y^2_{\!_{\phi \chi}} \frac{ m_{\phi, \rm eff}}{8 \pi} \left( 1- \frac{4 m_{\chi, \rm eff}^2}{ m_{\phi, \rm eff}^2}\right)^{3/2}
\, ,
\end{eqnarray}
where $c_\psi$ denotes the color factor, and
$m_{\chi, \rm eff}   \simeq    \vert  m_\chi + y_{\!_{\phi \chi}} \phi  \vert$ and $m_{\psi, \rm eff}   \simeq   \vert  m_\psi^{\rm th} + y_{\!_{\phi \pi}} \phi \vert$ are the effective masses of DM and RH SM fermions, respectively, where $m_\psi^{\rm th}$ denotes the plasma induced thermal masses for the RH SM sector~\citep{Weldon:1982bn}.
For $m_{\chi, \rm eff}, m_{\psi, \rm eff} \ll m_{\phi, \rm eff}$ and $\phi \ll M_P$, we can approximate $\Gamma_\phi \simeq \alpha_\phi m_{\phi, \rm eff} \simeq \alpha_\phi M/\sqrt{\alpha}$.
Assuming universal inflaton coupling to RH SM fermions and suppressing the coupling to RH neutrinos (if any), we have $\alpha_\phi \simeq 21 \ y_{\!_{\phi \psi}}^2/8\pi$, where $\alpha_\phi$ can be defined in terms of the reheat temperature $T_{\rm rh}$ through the relation~\citep{Kolb:1990vq,Chung:1998ua,Chung:1998rq} 
\begin{align}
H^2(\tau_\phi) \ \simeq \ \frac{\Gamma_\phi^{2}}{4} \ \simeq \  \frac{(\alpha_\phi m_\phi)^2}{4} \ \simeq \ \frac{1}{3 M_P^2}\frac{ \pi^2}{30} g_\rho T_{\rm rh}^4 \, ,
\end{align}
which gives  the following expression for $ \alpha_\phi $: 
\begin{eqnarray} \label{eq:alpha}
\alpha_\phi \ = \ \left( \frac{2 \pi^2}{45}  \right)^{1/2} g_\rho^{1/2} \frac{T_{\rm rh}^2}{M_P m_\phi} \, .
\end{eqnarray}
The inflaton branching ratio to DM is then given by
\begin{eqnarray} \label{eq:Bchi}
B_\chi \ \equiv \ \Gamma(\phi \rightarrow \overline{\chi} \chi)/\Gamma_\phi
\sim y^2_{\!_{\phi \chi}}/(21 \ y^2_{\!_{\phi \psi}}) \ll 1 
%= \  \frac{1}{8 \pi} \frac{y_{\!_{\phi \chi}}^2}{\alpha_\phi} \ \sim \ \frac{y_{\!_{\phi \chi}}^2}{21 y_{\!_{\phi \psi}}^2} \ \ll \ 1
\,.
\end{eqnarray}
Note that the finite temperature effects on the 
inflaton decay rate are negligible as long as the maximum attainable temperature, $ T_{\rm max} \ll m_\phi $, where~\citep{Kolb:1990vq,Chung:1998rq,Mazumdar:2013gya,Drewes:2014pfa}\footnote{In the derivation of Eq.~\eqref{eq:tmax}, the masses of daughter particles are assumed to be much smaller than that of inflaton.}
\begin{align}
T_{\rm max} \ \simeq \ \frac{1.64}{\sqrt{\pi}} g_\rho^{-1/4}  \Gamma_\phi^{1/4} M_P^{1/4} \rho_{\phi,I}^{1/8}, 
\label{eq:tmax}
\end{align}
with $ g_\rho $ being the number of relativistic d.o.f contributing to the radiation energy density and $\rho_{\phi,I} \simeq V_{\rm end}$
is the inflaton energy density at the end of inflation. As for the  thermal contribution, one can easily show that the relevant contribution comes from the $X$-mediated DM production in Fig.~\ref{fig:annihilation}. In particular, 
$\overline{\psi}_{R} \psi_R \rightarrow \phi \rightarrow \overline{\chi} \chi$ would yield a sub-dominant DM contribution compared to $\overline{\psi}_{R} \psi_R \rightarrow X \rightarrow \overline{\chi} \chi $. This is due to the fact that both thermal and non-thermal contribution to DM abundance in the inflaton-mediation case are proportional to $ y_{\!_{\phi \chi}} $ [cf. Eq.~(\ref{eq:Bchi})], which has to be small in order not to non-thermally overproduce DM~\citep{Allahverdi:2002nb,Dev:2013yza}. 
The thermal contribution due to inflaton mediation dominates only when $ T_{\rm max} \gg m_\phi $, in which case the thermal corrections to inflaton decay rate also become important~\citep{Drewes:2014pfa}. On the other hand, such complications do not arise in case of the $X$-mediation as long as $m_X\gtrsim T_{\rm max}$.

Assuming $ B_\chi \ll 1 $, we can trace the time evolution of the inflaton decay products by solving the following  set of coupled
Boltzmann equations~\citep{Kolb:1990vq,Rychkov:2007uq}\footnote{This set of Boltzmann equations is valid only when $\phi$ is oscillating around a quadratic minimum, which is quickly realized in our case only after a few oscillations, when $\phi\ll M_P$.  % One can instead multiply Eq.~(\ref{eq:phi}) by $\dot{\phi}$ and use the fact that $\rho_\phi+P_\phi = \dot{\phi}^2$ for the $\phi$ homogeneous condensate to obtain
%
%$$\dot{\rho}_\phi \: + \: 3 H (\rho_{\phi}+P_{\phi})  \  = \  -\Gamma_{\phi} \dot{\phi}^2 \, . $$
%
%One can then use conservation of energy to derive the relevant Boltzmann equations for the daughter particles.
%However, we numerically verified that after a few $\phi$ oscillation the energy and number densities coincide and that the DM abundances are exactly the same.
}:
\begin{eqnarray}
\label{eq:rhophi}
&& \frac{d{\rho}_\phi}{dt}  \: + \: 3 H \rho_{\phi}  \  = \  -\Gamma_{\phi} \rho_{\phi}\\
\label{eq:rhorad}
&& \frac{d{\rho}_{\rm rad}}{dt} \: + \: 4 H \rho_{\rm rad} \  = \ (1-B_{\chi}) \Gamma_{\phi} \rho_{\phi} \\
\label{eq:nchi}
%\begin{split}
&& \frac{d{n}_\chi}{dt} \: + \: 3 H n_{\chi} \  = \  2 B_{\chi} \Gamma_{\phi}\frac{\rho_{\phi}}{m_\phi} \: + \: \sum_{\psi} \gamma(\overline{\psi}_R \psi_R \rightarrow \overline{\chi} \chi) \,,
%\end{split}
\end{eqnarray}
where $ \rho_\phi $  ($ \rho_{\rm rad} $) denotes the inflaton (radiation) energy density, $n_\chi$ is the DM number density, $\gamma$ is the DM thermal production rate and $m_\phi$ is the effective inflaton mass averaged over one oscillation, as shown in Fig.~\ref{fig:mphi-r}. 
For simplicity, we focus on the case of heavy mediator i.e. $ m_X \gg T_{\rm max} $. Then on dimensional grounds, the cross section $\sigma(\overline{\psi}_{R}\psi_R \rightarrow \overline{\chi} \chi) \sim y_{\!_{X\chi}}^2 y_{\!_{X \psi}}^2 ~ T^2/m_X^4$, and since the DM thermal production rate  $ \gamma \propto \sigma$, we have 
\begin{eqnarray} \label{eq:gamma}
\gamma(\overline{\psi}_{R} \psi_R \rightarrow \overline{\chi} \chi) \ = \ \mathcal{I}(\overline{\psi}_{R} \psi_R \rightarrow \overline{\chi} \chi) \ \frac{T^8}{m_X^4}\,,
\end{eqnarray}
where $ \mathcal{I} \sim y_{\!_{X \chi}}^2 y_{\!_{X \psi}}^2 $ is a constant for $ T \gg m_{\chi, \rm eff}, m_{\psi, \rm eff} $ and the factor $T^{6}$ arises from the two $ \psi_R$'s being in a thermal plasma. In the numerical calculations we use the exact integral expression for $ \gamma$~\citep{Gondolo:1990dk} and also take into account the plasma induced thermal masses for RH SM fermions~\citep{Weldon:1982bn}. %,  see Appendix.

%%%%%%%%%%%%%%%%%%%%%%%%%%%%%%%%%%%%%%%%%%%%%%%%%%%%%%%%%%%%%%%%%%%%%%%%%%%%%%%%%%%%%%%%%%%%%%%%%%%%%%%%%%%%%%%%%%%%%%%%%%%%%%
\section{THERMAL AND NON-THERMAL DARK MATTER ABUNDANCE}
\indent

The total DM relic abundance for temperatures $T \ll T_{\rm rh}$ is roughly given by the sum of the thermal (th) and non-thermal (non-th) components~\citep{Gelmini:2006pw,Gelmini:2006pq}: 
\begin{eqnarray} \label{eq:omega}
\Omega_\chi h^2 \  \simeq  \ \Omega_\chi^{\rm non-th} h^2 + \Omega_\chi^{\rm th} h^2 
  \   \simeq  \ 2.74 \times 10^8 \ (n_{\chi}^{\rm non-th} + n_{\chi}^{\rm th})\frac{m_\chi}{s} \,,
\end{eqnarray}
where $h$ is the scaled Hubble rate, $ s = (2\pi^2/45) g_s T^3 $ is the entropy density with $ g_s $ being the corresponding number of relativistic d.o.f. We take $g_\rho=g_s \equiv g$, which is valid for most of the thermal history of the Universe. Eqs.~(\ref{eq:rhophi})-(\ref{eq:nchi}) can be simplified by defining %dimensionless
comoving energy and number densities~\citep{Chung:1998rq}:  
%$ \Phi \equiv \rho_\phi x^3/m_\phi^4 $, $ R \equiv \rho_{\rm rad} x^4/m_\phi^4 $ and $ X \equiv n_\chi x^3/m_\phi^3 $, with $ x(t) \equiv a(t) m_\phi $ where $ a(t) $ is the scale factor.
$ \Phi \equiv \rho_\phi a^3$, $ R \equiv \rho_{\rm rad} a^4$ and $ X \equiv n_\chi a^3$ where $a$ is the scale factor. The temperature of the plasma during inflaton oscillations reaches its maximum value  $T_{\rm max}$ given by Eq.~\eqref{eq:tmax} at roughly $a_{\rm max}/a_{I} \sim 1.5$, where $a_I$ denotes the initial scale factor at the end of inflation. Thereafter, the temperature decreases with the expansion as $a^{-3/8}$
until radiation takes over signalling the end of reheating phase at $a_{\rm rh}$. Similarly,
the DM thermal and non-thermal abundances, $Y_\chi^{\rm th(non-th)} \equiv n_\chi^{\rm th(non-th)} / s$, increase initially very fast to reach their maxima very soon after the beginning of the reheating process and then decrease as $a^{-3/8}$ till the end of reheating. 

In the non-thermal case [Fig.~\ref{fig:decay}], the DM relic abundance is directly determined from the inflaton energy density, which it turn depends on $r(\alpha)$ [see the first term on the RHS of Eq.~(\ref{eq:nchi}) and Fig.~\ref{fig:Vend-r}], besides the sensitivity of the decay rate to the shape of the inflationary potential around the minimum which also depends on $r(\alpha)$. Thus, the connection between the DM properties and the primordial fluctuations is straightforward, but to our knowledge, this important point was never discussed in the literature. We further show that such an interesting connection also exists in case (b) [see Fig.~\ref{fig:annihilation}] if the mediator mass is between $T_{\rm max}$ and $M_P$. 
This can be understood from Eqs.~(\ref{eq:nchi}) and (\ref{eq:gamma}), where the term sourcing the DM thermal abundance is proportional to $T^8 \sim \rho_{\rm rad}^{2}$ and since the thermal bath itself arises from the decay of inflaton [cf. Eq.~(\ref{eq:rhorad})], a connection between DM thermal abundance and the tensor-to-scalar ratio can be established.

%However, there are two regimes of interest. When $\alpha \gg 1$, then $m_\phi$ becomes constant and the inflationary potential mimics that of the quadratic potential, see Fig.~(\ref{fig:pot}). In this case, the correlation between $\Omega_{\chi}h^2$ and $r$ will be very mild. This case has been briefly discussed in an Appendix, analytically for both thermal and non-thermal production of DM, see Appendix.

%When $\alpha \sim {\cal O}(1)$, in this case the inflaton potential around the minimum is not quadratic, see Fig.~(\ref{fig:pot}), and it will be interesting to see how to correlate $\Omega_{\chi}h^2$ with respect to tensor-to-scalar ratio $r$.

\begin{figure}[t] 
\centering
\subfigure[]{\label{fig:omega_th-r}
\includegraphics[width=6.0cm]{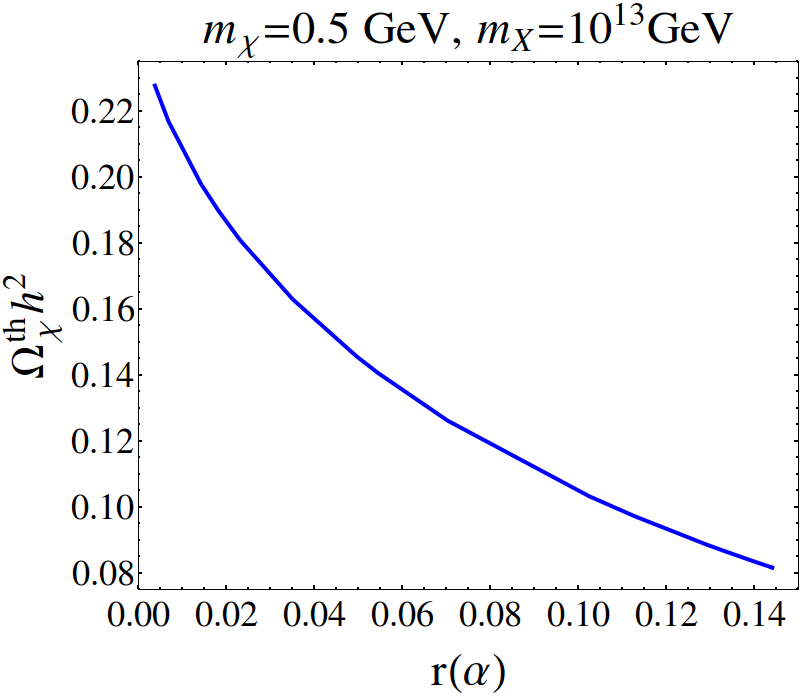}}
\hspace{0cm}
\subfigure[]{\label{fig:omega_nonth-r}
\includegraphics[width=6.0cm]{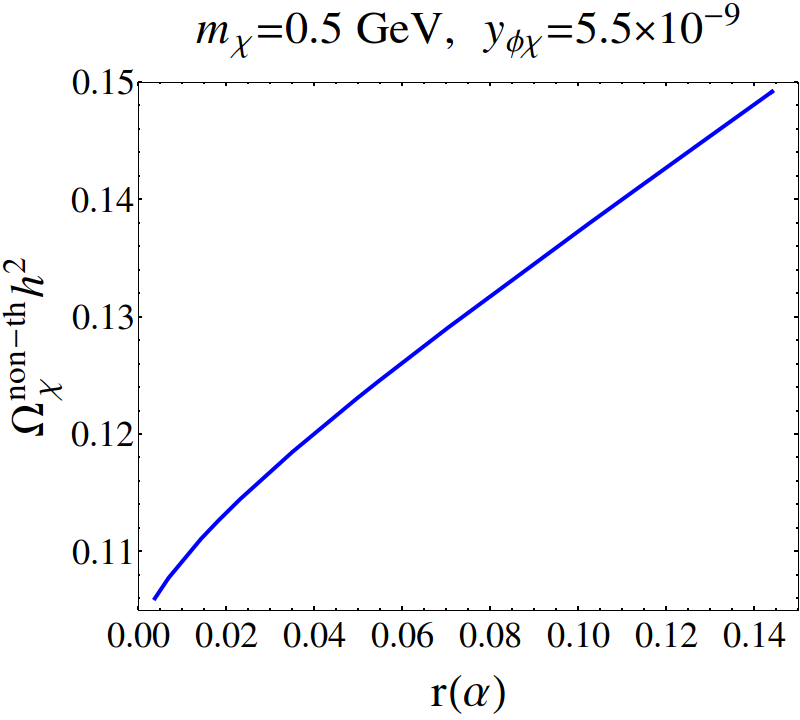}}
%\subfigure[]{\label{fig:Trh-r}
%\includegraphics[width=6.0cm]{Trh-r.png}}
\caption{Dark matter (a) thermal and (b) non-thermal abundance as a function of the tensor-to-scalar ratio. Here we fix $m_\chi=0.5$~GeV, $y_{\!_{X \chi}}=y_{\!_{X \psi}}=1$ and $y_{\!_{\phi \psi}}=3.6 \times 10^{-7}$. 
% which results in a reheating temperature, $T_{\rm rh} \simeq 10^{9}$~GeV for $r \simeq 0.004$ ($\alpha=1$). For larger $r$, $T_{\rm rh}$ is less.
}
\label{fig:omega-r}
\end{figure}

%%%%%%%%%%%%%%%%%%%%%%%%%%%%%%%%%%%%%%%%%%%%%%%%%%%%%%%%%%%%%%%%%%%%%%%%%%%%%%%%%%%%%%%%%%%%%%%%%%%%%%%%%%%%%%%%%%%%%%%%%%%%%%
%%\indent
In the Appendix, we have obtained approximate analytical expressions for the DM thermal and non-thermal abundances, and argued that in both cases, the details of the inflationary potential are carried over to DM abundance via both $m_\phi$ and $V_{\rm end}$, thereby establishing a connection between the DM abundance and tensor-to-scalar ratio. 
%Figs.~\ref{fig:Vend-r} and \ref{fig:mphi-r} respectively show the dependence of $V_{\rm end}$ and $m_\phi$ on $r(\alpha)$. 
In order to precisely capture this connection between $r$ and $\Omega_\chi h^2$, we  need to integrate Eqs.~(\ref{eq:rhophi})-(\ref{eq:nchi}) numerically for different $r(\alpha)$ values. The resulting  DM thermal and non-thermal abundances are shown in Figs.~\ref{fig:omega-r} (a) and (b), respectively, for a typical choice of parameters: $m_\chi=0.5$~GeV, $y_{\!_{X \chi}}=y_{\!_{X \psi}}=1$ and $y_{\!_{\phi \psi}}=3.6 \times 10^{-7}$. 
We find that for the thermal case, $\Omega_\chi^{\rm th}\propto r^{-3/8}$, while for the non-thermal case,  $\Omega_\chi^{\rm non-th}\propto r^{4/5}$. This is a very interesting connection, especially the former one, since it relates thermal production of DM at the time of reheating and thermalization
of the Universe with the value of $r$ which,  as a matter of fact, 
provides a promising probe for the FIMP-like scenario in near future. In other words, if the tensor-to-scalar ratio is measured in future, one can test the properties of FIMP DM through this connection in the context of a given inflationary model.
%One can easily see that the DM abundance either produced thermally or non-thermally depends on the tensor-to-scalar ratio, $r$. For the non-thermal case, the DM abundance is proportional to $r$, whereas the abundance of the thermally produced DM decreases as $r$ increases.
%
%%%%%%%%%%%%%%%%%%%%%%%%%%%%%%%%%%%%%%%%%%%%%%%%%%%%%%%%%%%%%%%%%%%%%%%%%%%%%%%%%%%%%%%%%%%%
%%%%%%%%%%%%%%%%%%%%%%%%%%%%%%%%%%%%%%%%%%%%%%%%%%%%%%%%%%%%%%%%%%%%%%%%%%%%%%%%%%%%%%%%%%%%
%
%

\begin{figure}[t] 
\centering
\includegraphics[width=7.0cm]{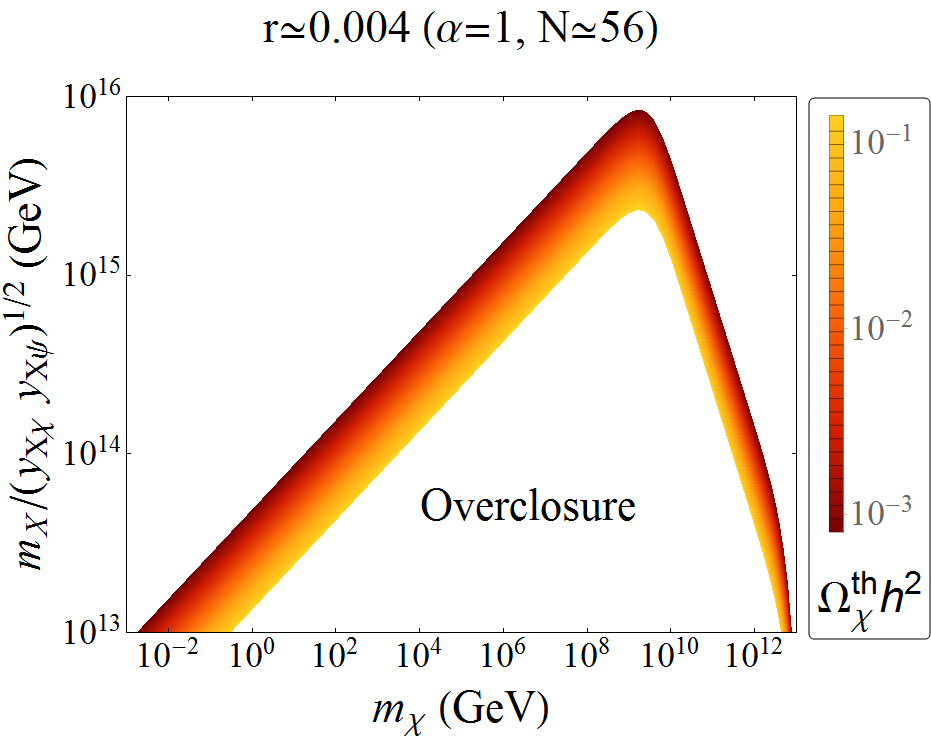}
%\hspace{0.0 cm}
%\subfigure[]{\label{fig:scan1000}
%\includegraphics[width=6.0cm]{scan1000.png}}
\caption{Map of the DM thermal abundance as a function of $ m_\chi $ and $ m_X$ for $r \simeq 0.004$ and $T_{\rm rh} \simeq 10^{9}$~GeV. 
%The unshaded region labelled by overclosure is ruled out by overclosure constraints on Dark matter abundance, while the rest of the unshaded region is still allowed, though for practical purposes, the corresponding DM thermal abundance becomes negligible.}
}
\label{fig:scan}
\end{figure}

To examine the dependence on the mediator mass,  we scan the $(m_\chi, m_X)$ parameter space for the correct DM abundance while fixing the mediator couplings to DM and the SM d.o.f.
In Fig.~\ref{fig:scan}, we show the DM thermal abundance heat map as a function of $ m_\chi $ and $ m_X $ for 
$r \simeq 0.004$. 
% (left panel) and $r \simeq 0.111$ (left panel).
We have fixed $y_{\!_{\phi \psi}} = 3.6 \times 10^{-7}$, so that  $T_{\rm rh} \simeq 10^{9}$~GeV.
%for $r \simeq 0.004$ and $ 7.5 \times 10^{8}$~GeV for $r \simeq 0.111$.
For larger $r$, the allowed range of ($m_\chi$, $m_X$) shifts to larger values of $m_\chi$ and smaller $m_X$ values.
Similarly, for smaller $y_{\!_{\phi \psi}}$, i.e. smaller reheating temperature, the allowed region of the parameter space shifts to larger DM masses and smaller mediator masses, and vice-versa.
For very small values of the branching fraction $B_\chi \ll 1$, the thermal contribution given by Eq.~(\ref{eq:omegath}) can be dominant over the non-thermal contribution given by Eq.~(\ref{eq:omeganth}), and can account for the observed DM abundance for $ m_\chi $ as low as roughly 300 MeV/$(y_{\!_{X \chi}} y_{\!_{X \psi}})^{2}$.
The unshaded region labelled by overclosure (below the colored region) gives $\Omega_\chi h^2 > 0.13$, which is ruled out at $3\sigma$ by the latest Planck data~\citep{Ade:2015xua}. The rest of the unshaded region (above the colored region) is still allowed, though for practical purposes, 
the corresponding DM thermal abundance becomes  negligible, and one has to allow for a non-thermal contribution to the DM abundance or to invoke a multi-component DM to explain the observed abundance, see e.g.~\citep{Chialva:2012rq}.  
Note that for $ m_\chi \gtrsim T_{\rm rh} $, the thermal production rate is suppressed due to a smaller phase space. This can be seen from the right parts of the allowed ($m_\chi$, $m_X$) parameter space in Fig.~\ref{fig:scan}.
%Fixing the tensor-to-scalar ratio, higher values of $ T_{\rm rh} $ causes the allowed range of ($m_\chi$, $m_X$) to shift to lower values of $ m_\chi$ and vice-versa.
%
%%%%%%%%%%%%%%%%%%%%%%%%%%%%%%%%%%%%%%%%%%%%%%%%%%%%%%%%%%%%%%%%%%%%%%%%%%%%%%%%%%%%%%%%%%%%
%%%%%%%%%%%%%%%%%%%%%%%%%%%%%%%%%%%%%%%%%%%%%%%%%%%%%%%%%%%%%%%%%%%%%%%%%%%%%%%%%%%%%%%%%%%%
%

%%%%%%%%%%%%%%%%%%%%%%%%%%%%%%%%%%%%%%%%%%%%%%%%%%%%%%%%%%%%%%%%%%%%%%%%%%%%%%%%%%%%%%%%%%%%%%%%%%%%%%%%%%%%%%%%%%%%%%%%%%%%%%
%%%%%%%%%%%%%%%%%%%%%%%%%%%%%%%%%%%%%%%%%%%%%%%%%%%%%%%%%%%%%%%%%%%%%%%%%%%%%%%%%%%%%%%%%%%%%%%%%%%%%%%%%%%%%%%%%%%%%%%%%%%%%%
\section{DISCUSSION AND CONCLUSION}
\indent

The  BICEP2 collaboration had recently claimed a positive detection of the primordial tensor modes with more than $5\sigma$ significance~\citep{Ade:2014xna}. However, a more careful analysis of the foreground dust weakened the claim to an upper limit of $r<0.12$ at 95\% CL~\citep{Ade:2015tva}. Nevertheless, if the tensor-to-scalar ratio is definitively measured in future, it will provide an important  indirect handle on the dark sector physics, as shown above, along with a clear evidence of a quantum nature of gravity~\citep{Ashoorioon:2012kh}. 

Before we conclude, let us briefly mention that we could have also imagined a similar mechanism for producing baryon/lepton (B/L) asymmetry, either from the direct decay of the 
inflaton, if the inflaton were carrying any B/L number~\citep{Murayama:1992ua}, or from the intermediate condensate $X$ carrying B/L number~\citep{Enqvist:2003gh}, as e.g.    
in GUT-baryogenesis~\citep{Kolb:1990vq}. As a concrete example, in non-supersymmetric models, we can realize high-scale thermal leptogenesis via the production and decay of RH heavy Majorana neutrinos~\citep{Davidson:2008bu}. In either case, we would be able to relate the scale of inflation, and therefore the tensor-to-scalar ratio, with the magnitude of the L-asymmetry. In the latter case, one would require a weak washout regime in order to retain the sensitivity towards the initial conditions. A detailed discussion of these issues will be given elsewhere. 

To conclude, if the future observations could pin down the exact value of the tensor-to-scalar ratio, it would serve as an interesting way to constrain the hidden sector, including the properties of the DM feebly interacting with the SM d.o.f, which are otherwise very hard to probe. Although for the sake of illustration we have used a particular class of inflationary potential to derive our results,
the idea of connecting the DM abundance to the primordial tensor perturbations should hold true for a generic model of inflation, including multi field driven inflationary models~\citep{Liddle:1998jc}.

\section*{ACKNOWLEDGMENTS}
\indent

The work  of P.S.B.D. and   A.M. are   supported   by  the  STFC grant ST/L000520/1.
S.Q. is funded by the King Abdulaziz University.
A.M. would like to thank the
hospitality of Ruth Durrer and the Universit\'{e} de Gen\'{e}ve where
part of the work was completed. P.S.B.D. thanks the local hospitality at IIT, Guwahati, during the last phase of the work. 
%%%%%%%%%%%%%%%%%

%%%%%%%%%%%%%%%%%%%%%%%%%%%%%%%%%%%%%%%%%%%%%%%%%%%%%%%%%%%%%%%%%%%%%%%%%%%%%%%%%%%%%%%%%%%%%%%%%%%%%%%%%%%%%%%%%%%%%%%%%%%%%%
\section*{APPENDIX: ANALYTICAL ESTIMATION OF THE DM ABUNDANCE}
\indent

With the assumptions mentioned in Section 3.1, namely, (i) $m_\phi \ll M_P$, (ii)  $y_{\!_{\phi \psi}}, y_{\!_{\phi \chi}} \ll m_\phi/M_P$, and (iii) $m_{\psi}^{\rm th}, m_{\chi} \ll m_{\phi}$, we derive analytical expressions for the DM abundance produced either thermally or non-thermally.

\noindent
{\bf Non-thermal abundance:}
Deep inside the inflaton domination (id) epoch, i.e when $ H \gg \Gamma_\phi $, $ \Phi \simeq \Phi_I$ where $ \Phi_I = \rho_{\phi,I} a_I^3 \simeq V_{\rm end} a_I^3$ and $ R_I \simeq X_I \simeq 0 $; so  Eq.~(\ref{eq:rhorad}) can be easily solved to yield 
\begin{eqnarray}
%\rho_{\rm rad}^{\rm id }(a) \ & \simeq & \ \frac{2\sqrt{3}}{5} \Gamma_\phi M_P  \rho_{\phi,I}^{1/2} \left(\frac{a_I}{a}\right)^{3/2}
\rho_{\rm rad}^{\rm id }(a) \  \simeq  \ \frac{2\sqrt{3}}{5} \Gamma_\phi M_P  V_{\rm end}^{1/2} \left[ \left(\frac{a_I}{a}\right)^{3/2} - \left(\frac{a_I}{a}\right)^{4} \right]
\, .
\end{eqnarray}
This allows us to estimate  the temperature of the ambient relativistic d.o.f during the inflaton domination epoch~\citep{Chung:1998rq}:
\begin{eqnarray} \label{eq:Tid}
T^{\rm id}(a)  \ = \ \left(  \frac{30}{\pi^2 g_{\rho}}   \rho_{\rm rad}^{\rm id}\right)^{1/4}  \simeq  \left( \frac{432}{\pi^4 g_\rho^2}\right)^{1/8} \Gamma_\phi^{1/4} M_P^{1/4}  V_{\rm end}^{1/8} \left[ \left(\frac{a_I}{a}\right)^{3/2} - \left(\frac{a_I}{a}\right)^{4}  \right]^{1/4}
 .
\end{eqnarray}
Now Eq.~(\ref{eq:nchi}) can be easily solved for the DM non-thermal component, which is sourced by the first term on the RHS of Eq.~(\ref{eq:nchi}).
%, to yield
%
%
%\begin{align}
%n_\chi^{\rm id,non-th}(a) \ \simeq \ \frac{4}{\sqrt{3}} B_\chi \alpha_\phi M_P \rho_{\phi,I}^{1/2} \left(\frac{a_I}{a}\right)^{3/2}
%n_\chi^{\rm id,non-th}(a) \ \simeq \ \frac{4}{\sqrt{3}} B_\chi \alpha_\phi M_P V_{\rm end}^{1/2} \left[\left(\frac{a_I}{a}\right)^{3/2} - \left(\frac{a_I}{a}\right)^{3}   \right]
%\, .
%\end{align} 
%
%
Evaluating this general expression for the non-thermal number density at $ T_{\rm rh} $,  and accounting for the inflaton population decaying at $ T<T_{\rm rh} $ and the accompanying entropy release~\citep{Giudice:2000ex}, we reproduce the non-thermal contribution to the relic density~\citep{Dev:2013yza, Allahverdi:2002nb}
\begin{align} \label{eq:omeganth}
\frac{\Omega_\chi^{\rm non-th} h^2}{0.12} \ \simeq  \   & 
3.87 \times 10^5 \ B_\chi  \left(\frac{g}{106.75} \right)^{-1/4}  \left( \frac{\alpha_\phi}{10^{-13}}\right)^{1/2} \left(\frac{m_\chi}{1~{\rm GeV}}\right)  \left( \frac{m_{\phi}}{10^{13}~{\rm GeV}}\right)^{-3/4} 
\left( \frac{V_{\rm end}}{(10^{16}~{\rm GeV})^4} \right)^{1/8}
\nonumber \\ 
\ \simeq \ & 4.44 \times 10^5 \ B_\chi  \left(\frac{m_\chi}{1~{\rm GeV}}\right) \left( \frac{m_{\phi}}{10^{13}~{\rm GeV}}\right)^{-1} \left( \frac{T_{\rm rh}}{10^{9}~{\rm GeV}}\right)
 \, .
\end{align} 
Here we have used Eq.~(\ref{eq:alpha}) for $ \alpha_\phi$ and Eq~(\ref{eq:Tid}) to change the dependence from $ a_{\rm rh} $ to $ T_{\rm rh} $.
Clearly, the DM non-thermal abundance depends on the inflaton energy at the end of inflation, $V_{\rm end}$, and it is also sensitive to the steepness of the inflationary potential around the minimum characterized by $m_\phi$. In general, both $V_{\rm end}$ and $m_\phi$ depend on $r(\alpha)$ [cf. Fig.~\ref{fig:Vend-mphi}] for the class of potentials under consideration. This gives rise to the $\Omega_\chi^{\rm non-th} \propto r^{4/5}$ behavior in Fig.~\ref{fig:omega_nonth-r}. 

%Form Fig.~\ref{fig:r} one can see however that this dependence becomes very mild when $\alpha \gtrsim 10^4$.

%~\footnote{ Note that in order to avoid the overclosure of the Universe due to the DM non-thermal production from the inflaton, one typically requires a small branching fraction, which is a challenge for many models of inflation and non-thermal production of DM, in particular string compactification models~\citep{cicoli, Chialva:2012rq}, due to the proliferation of hidden sectors. We assumed that the $X$ field is not excited after inflation, since it is kinematically forbidden if $m_X > m_\phi/2$, and resonant production is suppressed due to small couplings between $\phi$ and $X$. Thus, there is no non-thermal contribution to DM abundance from $X$ decay.}.

\noindent
{\bf Thermal contribution:}
For thermal contribution, since $ \gamma \propto T^8 $ [cf. Eq.~(\ref{eq:gamma})], which during the radiation domination epoch will redshift as $a^{-8}$, the scaled version of Eq.~(\ref{eq:nchi}) will in turn go as $d[n_\chi^{\rm th} a^3]/da \propto a^{-4}$. This means that except for a dilution factor due to the entropy released directly after the transition to the radiation domination phase,
%at $ t \gtrsim \Gamma_\phi^{-1}$, 
the DM thermal yield becomes constant directly after the end of reheating $Y_\chi^{\rm th} (T \ll T_{\rm rh}) \simeq \zeta \ Y_\chi^{\rm th} (T = T_{\rm rh})$, where $\zeta$ is the dilution factor. Thus, we obtain 
%Therefore, for the DM thermal abundance, it is sufficient to evaluate the yield function $ n_\chi^{\rm th}/s $ at $ T_{\rm rh} $, and account for the entropy released after reheating. Substituting Eqs.~(\ref{eq:omega}) and (\ref{eq:Tid}) into Eq.~(\ref{eq:nchi}), we obtain
%
%\begin{eqnarray}
%n_\chi^{\rm id,th} (a) \ &\simeq & \ \frac{20736 \sqrt{3}}{\pi^9 g^2} y_{\!_{X \chi}}^2 y_{\!_{X \psi}}^2 \Gamma_\phi^2 M_P^3  m_X^{-4} V_{\rm end}^{1/2} \left[ 7 \left(\frac{a_I}{a}\right)^{3/2} - 25 \left(\frac{a_I}{a}\right)^{3} + 21 \left(\frac{a_I}{a}\right)^{4} \right.
%\nonumber \\
%&& \left. - 3 \left(\frac{a_I}{a}\right)^{13/2}\right] \, .
%\label{eq:nchith}
%\end{eqnarray}
%
%The evolution of the thermal yield $ Y_\chi^{\rm th} $ with respect to the scale factor  $ a $, normalized to $ a_I $.The abundance $ Y_\chi^{\rm th} $ builds up very quickly as the thermal scattering rate $ \gamma $ increases with temperature [cf. Eq.~(\ref{eq:gamma})], to reach a maximum very soon after the beginning of the reheating process then decreases as $a^{-3/8}$ 
%since $s \sim T^3 \propto a^{-9/8}$, $Y_\chi^{\rm th}= n_\chi^{\rm th}/s$, due to the ongoing entropy production from  the inflaton decay, roughly till the end of reheating phase at $a_{\rm rh}$ where $\chi$ completely decouples from the thermal bath. %Fig.~\ref{fig:DM_decoupling} also shows the evolution of the scaled energy densities for inflaton ($ \Phi $) and radiation ($ R $) for comparison.
% with $ \mathcal{I} \simeq (126/\pi^5)  y_{\!_{X \chi}}^2 y_{\!_{X \psi}}^2  $, 
%From Eq.~(\ref{eq:nchith}) 
%
\begin{align} \label{eq:omegath}
\frac{\Omega_\chi^{\rm th} h^2}{0.12}  
  \ \simeq \  & 
1.73 \ y_{\!_{X \chi}}^2 y_{\!_{X \psi}}^2  \left(\frac{g}{106.75} \right)^{-9/4}  \left( \frac{\alpha_\phi}{10^{-13}}\right)^{3/2}  \left(\frac{m_\chi}{1~{\rm GeV}}\right)  \left( \frac{m_X}{10^{13}~{\rm GeV}}\right)^{-4} 
%\nonumber \\
%& 
\left( \frac{m_{\phi, \rm eff}}{10^{13}~{\rm GeV}} \right)^{3/4} \left( \frac{V_{\rm end}}{(10^{16}~{\rm GeV})^4} \right)^{3/8}
\nonumber \\
 \ \simeq \ & 
2.61  \ y_{\!_{X \chi}}^2 y_{\!_{X \psi}}^2  \left(\frac{g}{106.75}\right)^{-3/2} \left(\frac{m_\chi}{1~{\rm GeV}}\right)  \left( \frac{m_X}{10^{13}~{\rm GeV}} \right)^{-4} \left( \frac{T_{\rm rh}}{10^{9}~{\rm GeV}} \right)^3 
\, .                                  
\end{align}
Here again we have used Eq.~(\ref{eq:alpha}) for $ \alpha_\phi$ and Eq~(\ref{eq:Tid}) to change the dependence from $ a_{\rm rh} $ to $ T_{\rm rh} $.
It is clear that the DM thermal abundance is also sensitive to the inflaton energy at the end of inflation and inflaton mass around the minimum of the potential, see Figs.~\ref{fig:Vend-r} and \ref{fig:mphi-r}. Hence follows the $\Omega_\chi^{\rm th} \propto r^{-3/8}$ behavior in Fig.~\ref{fig:omega_th-r}.

\bibliographystyle{apalike_p}
{
\fontsize{10}{10}\selectfont
\renewcommand\refname{REFERENCES}
\setlength{\bibhang}{1cm}
\setlength{\bibsep}{0.13cm}
\bibliography{FIMPv3_SCC9.bib}
}

%\fontsize{10}{10}\selectfont

\end{document}